\documentclass[11pt,english]{article}
\usepackage{geometry}
\usepackage{amsthm}
\usepackage{amsmath}
\usepackage{amssymb}
\usepackage{array}
\usepackage{enumerate}
\usepackage{graphicx}
\usepackage{setspace}
\usepackage{tabu}
\usepackage{tabularx}
\usepackage{color}
\usepackage{braket}
\usepackage{enumerate}	
\usepackage[table,xcdraw]{xcolor}
\usepackage{titling}
\usepackage[affil-it]{authblk}
\usepackage[numbers]{natbib} 
\usepackage[hyphens]{url}
\usepackage{hyperref}
\usepackage[hyphenbreaks]{breakurl}
\begin{document}
	\title{Institutionalising Ethics in AI through Broader Impact Requirements}
	\author[1,2]{Carina E. A. Prunkl\thanks{\texttt{
				carina.prunkl@philosophy.ox.ac.uk}}
	} 
	\author[2]{Carolyn Ashurst}
	\author[2]{Markus Anderljung}
	\author[3]{Helena Webb}
	\author[2]{Jan Leike}
	\author[2]{Allan Dafoe}
	\affil[1]{Institute for Ethics in AI, University of Oxford}
	\affil[2]{Future of Humanity Institute, University of Oxford}
	\affil[3]{Department of Computer Science, University of Oxford} 
	
	\date{17 February 2021}
	\maketitle
	{\small \begin{abstract}
			Turning principles into practice is one of the most pressing challenges of artificial intelligence (AI) governance. In this article, we reflect on a novel governance initiative by one of the world's largest AI conferences. In 2020, the Conference on Neural Information Processing Systems (NeurIPS) introduced a requirement for submitting authors to include a statement on the broader societal impacts of their research. Drawing insights from similar governance initiatives, including institutional review boards (IRBs) and impact requirements for funding applications, we investigate the risks, challenges and potential benefits of such an initiative. Among the challenges, we list a lack of recognised best practice and procedural transparency, researcher opportunity costs, institutional and social pressures, cognitive biases, and the inherently difficult nature of the task. The potential benefits, on the other hand, include improved anticipation and identification of impacts, better communication with policy and governance experts, and a general strengthening of the norms around responsible research. To maximise the chance of success, we recommend measures to increase transparency, improve guidance, create incentives to engage earnestly with the process, and facilitate public deliberation on the requirement's merits and future. Perhaps the most important contribution from this analysis are the insights we can gain regarding effective community-based governance and the role and responsibility of the AI research community more broadly.
	\end{abstract}}
	
	\section*{Introduction}
	Growing concerns about the ethical, societal, environmental, and economic impacts of artificial intelligence (AI) have led to a wealth of governance initiatives. In addition to traditional regulatory approaches, complementary forms of governance can help address these challenges \citep{winfield_ethical_2018}. One such governance form is community-based technology governance, or ``governance from within'' \citep{fisher_midstream_2006}. Here, measures to influence research based on societal considerations develop from within the scientific community and are  implemented at the community level. A recent initiative of this kind comes from one of the world's largest AI conferences, the Conference on Neural Information Processing Systems (NeurIPS). In early 2020, the committee announced a new submission requirement: submitting authors must now include a statement that addresses the broader impacts of their research, including its ``ethical aspects and future societal consequences'' \citep{neurips_call_2020}. NeurIPS' new requirement has triggered mixed reactions by the AI research community, with discussions about its purpose and effectiveness emerging in social media and other venues \citep{johnson_neurips_2020}. While few contest that there exists a real need to identify and address ethical and societal challenges from AI, the diversity in reactions illustrates that there is little consensus on the right approach, nor on what the responsibilities of individual researchers or the research community (including conferences) should be in the process \citep{brundage_artificial_2016,hecht_its_2018}. It also highlights the need for further discussion on the purpose, implementation, and effects of NeurIPS' new requirement and similar governance measures.
	
	With this article we seek to contribute to the discussion on NeurIPS' new requirement, as well as on broader impact requirements in conference submissions more generally. We compare the new requirement to other established governance mechanisms and provide an analysis of its implications. Our goal is to (a) identify and make explicit the risks and challenges associated with the introduction of NeurIPS' broader impact requirement, (b) propose a set of measures to address these challenges, and (c) by doing so contribute to the discussion on this and related governance efforts in AI. While we take up several points that have already been made in favour of broader impact statements, most notably by Hecht et al. \citep{hecht_its_2018}, our article also emphasises and expands upon the relevant risks and challenges. In particular, our analysis shows how the success of NeurIPS' new impact statement requirement is by no means guaranteed---in fact, we believe that for the requirement to be effective, a substantial amount of effort must go into addressing the challenges listed below. Whether this effort suffices and even whether the requirement is warranted in the face of other, possibly more potent governance initiatives, will need to be subject of continued discussion. Despite this uncertain outcome of NeurIPS' unprecedented step, this initiative provides the community with a major opportunity to deliberate on what governance measures ought to be put in place so as to effectively address the challenges from AI development. 
	
	We begin our discussion by introducing NeurIPS' new requirement and comparing it to similar governance mechanisms. We then discuss the potential positive and negative effects, and identify several challenges and possible causes of these negative outcomes. Finally, we tentatively suggest several measures conference organisers and the research community at large can take so as to address the identified challenges.
	\section*{The new requirement and how it relates to existing practices}
	NeurIPS first announced their new impact statement requirement in a Medium post as follows:
	
	\begin{quote}
		\small Authors are asked to include a section in their submissions discussing the broader impact of their work, including possible societal consequences---both positive and negative \citep{neurips_call_2020}.
	\end{quote}
	
	In light of the community's increasing impact on society, the organisers write, it becomes necessary to ``think more broadly about what it means to develop new methods and systems, and to consider not only the beneficial applications and products enabled by our research, but also potential nefarious uses and the consequences of failure'' \citep{NeurIPS_getting_2020}. Papers are not rejected solely on the basis of the broader impact statement, though reviewers were asked to ``check whether the Broader Impact is adequately addressed'' \citep{neurips_faq_neurips_2020}. 
	
	In an analysis of the NeurIPS 2020 reviewing process, the program chairs provided more clarity on the process \citep{lin_what_2020}. Reviewers could flag submissions for ethical concerns and papers with strong technical reviews that had also been flagged for ethical concerns were passed to a pool of ethics advisers for assessment. Of the 13 papers forwarded, four were rejected and seven were conditionally accepted upon revision of the broader impact section or the removal of problematic data sets.
	
	NeurIPS' requirement to include a broader impact section in conference submissions is novel in many respects. Yet, similar measures exist across other disciplines and even within computer science. We briefly outline three related approaches.
	
	\subsection*{Institutional review boards (IRBs)}
	
	Research involving human subjects usually requires researchers to first obtain approval from their institutional review board (IRB). The purpose of IRBs is to assess whether the proposed research project meets certain ethical standards regarding the foreseeable impacts on human subjects. In addition to a project description, researchers have to submit an assessment of the potential risks associated with their project and suggest mitigating measures \citep{hamburger_new_2004}. Unlike the NeurIPS requirement, IRBs are less concerned with the broader impacts of research and its downstream applications, but instead focus on the direct effects on human subjects during or after the research process. Historically, IRBs have been rare in computer science, compared to the social and biomedical sciences. However, the IRB system and other ethical review processes are becoming more embedded in computer science. This development is in part driven by research conducted at the intersection of AI and disciplines that already have these processes in place, such as neuroscience, psychology, or sociology, and in part by a need for ethical oversight in the face of an increasing amount of empirical research involving human participants or their data \citep{buchanan_computer_2011,amorim_submit_2019}.
	
	There exists a large literature on the efficacy, structure, process, and outcome variance of IRBs \citep{abbott_systematic_2011}. Opinions on both the need for and the effectiveness of IRBs are mixed \citep{hyman_institutional_2007}, with critics lamenting excessive bureaucracy \citep{zywicki_institutional_2007}, lack of reliability \citep{whitney_principal_2008}, inefficiency \citep{chadwick_institutional_2000,fost_dysregulation_2007}, and, importantly, high variance in outcomes \citep{dziak_variations_2005,larson_survey_2004,shah_how_2004,mcwilliams_problematic_2003,goldman_inconsistency_1982}. Despite these difficulties, various surveys suggest that IRBs are generally perceived as a useful and necessary institution to protect human subjects, with proponents emphasising the need for ethical oversight to protect the public \citep{reeser_investigating_2008,stryjewski_impact_2015,keith-spiegel_what_2006,saleem_institutional_2011,chadwick_institutional_2000}. 
	
	\subsection*{Conference programme committees (PCs)}
	
	Some subfields within computer science, such as networking and cybersecurity, have implemented community-led oversight measures that go beyond IRB approval. For research involving users or user data, several conferences, including the Association for Computing Machinery's (ACM's) HotNets, SIGCOMM, IMC, and SIGMETRICS, not only require prior IRB approval, but also reserve the right to make their own assessment, in which they ``examine the ethical soundness of the paper'' \citep{acm_sigmetrics_call_2020}. In practice, rejections on the basis of ethical grounds are rare and opinions about the legitimacy and effectiveness of PC ethical oversight are mixed \citep{narayanan_no_2015,kenneally_cyber-security_2014}. One controversial case was a 2015 article on internet censorship, in which researchers collected information about website accessibility through the browsers of thousands of people across the world---without their knowledge or consent \citep{burnett_encore_2015}. After long debates, the PC of ACM SIGCOMM 2015 published the article but added a ``signing statement'' at the beginning of the article. There, they flagged the ethically controversial content. While this episode demonstrated that the PC was paying attention to the ethical implications of research, it also revealed that in cases where the community itself has not yet come to an agreement on relevant norms, the effectiveness of community-led ethical oversight can suffer as a result \citep{kenneally_cyber-security_2014}. A counterpoint to the above example is the publication of an article on ``emotional contagion'' through social networks by the journal PNAS \citep{kramer_experimental_2014}. Here, it was pressure from the community that led the journal to add an editorial expression of concern \citep{pnas_editorial_2014}.
	
	\subsection*{Funding bodies}
	
	Some funding bodies, including the US  National Science Foundation (NSF), UK Research and Innovation (UKRI) and the European Commission's Horizon2020 programme require grant applicants to embed indicators of impact in their applications, and some also require researchers to state how the eventual project will incorporate principles of Responsible Innovation \citep{epsrc_framework_2020}. To give an explicit example, the NSF states that it is interested in funding research related activities that lead to ``the advancement of scientific knowledge and activities that contribute to the achievement of societally relevant outcomes'' \citep{nsf_proposal_2018}. Next to ``intellectual merit'', ``broader impacts'' is one of two merit criteria for the assessment of grant applications.  Scientists have expressed mixed opinions about the criterion, with some embracing that it encourages scientists to reflect on the impacts of their work, while others consider it inappropriate, confusing, or burdensome and point out the low quality of many submitted statements \citep{tretkoff_nsfs_2007,frodeman_sciences_2007}. There are also debates on the broader question of whether peer-review, as is common practice in many funding agencies, is suitable for the assessment of broader impacts \citep{bozeman_broad_2009,holbrook_peer_2011,bozeman_socio-economic_2017,hecht_its_2018}. 
	
	While these requirements are an opportunity for grant applicants to highlight potential negative impacts alongside positive ones,  there is arguably little incentive for them to do so as the wording of funding bodies' application criteria tends to focus solely on the positive impacts through references to `public benefit' and `societal needs' etc. One case study in which researchers were explicitly asked about risks and ethical concerns comes from Owen and Goldberg \citep{owen_responsible_2010}. In this experiment, nanoscience researchers needed to  submit a risk register as part of an EPSRC grant application. They were asked to list ``any potential environmental, health, societal, or other impacts and/or any ethical concerns that may result from the innovation process'' (p.1701). Very few applicants addressed environmental impacts and none addressed future societal impacts. Most focused on relatively minor risks, such as laboratory safety. When interviewed, these applicants indicated that they lacked guidance on what impact assessment approaches to use (and how), which led to them feeling ``out of their depth'' (p.1704). Those who engaged more extensively with larger impacts had involved specialists from other disciplines in their project. 
	
	\subsection*{The NeurIPS broader impact requirement in comparison}
	
	There exist several differences between the above approaches and the NeurIPS broader impact requirement. First, unlike IRBs and the majority of the listed ACM conferences, the NeurIPS requirement is not restricted to research involving human subjects, users, or user data. All authors submitting to NeurIPS are required to include a broader impact section, though those working on very theoretical topics, may write ``that a Broader Impact discussion is not applicable'' \citep{neurips_faq_neurips_2020}. Second, the organisers require that the statement includes broader impacts, ethical aspects and ``future societal consequences'' \citep{neurips_call_2020}. Thus the statement is not limited to direct (or ``narrow'') impacts from the research process itself, contrary to both IRBs and the ACM conference examples. Third, NeurIPS requires authors to list both positive and \textit{negative} effects, which is similar to IRB applications but differs from many grant proposals. Finally, there are differences in timing. At the time of conference submission, research projects will have been completed and so any intervention will take place ex post, whereas interventions from IRBs and funding applications take place at a much earlier stage in the research cycle. 
	
	Table \ref{table1} summarises these similarities and differences between the different governance initiatives.
	
	\begin{table}[h!]
		\begin{center}
			\begin{tabularx}{\textwidth}{| r | >{\centering\arraybackslash}X | >{\centering\arraybackslash}X | >{\centering\arraybackslash}X | c |} \hline 
				& \textbf{Type of research} & \textbf{Impacts} & \textbf{Positive/ negative} & \textbf{Timing} \\ 
				\hline
				\textbf{NeurIPS requirement} & all* & broad & both & late \\ 
				\hline
				\textbf{IRBs} & human-subject & narrow & both & early \\ 
				\hline
				\textbf{ACM conferences} & varies & narrow & both & late \\
				\hline
				\textbf{Funding bodies} & all & broad & mainly positive & early \\
				\hline
			\end{tabularx}
			\caption{Comparison between the NeurIPS requirement and similar governance initiatives. By ``narrow'' impacts we refer to those resulting from the research process itself; by ``broad'' impacts we also include potential future consequences, for example those arising from future applications of the research. *Excluding very theoretical research.}
			\label{table1}
		\end{center}
	\end{table}
	
	We can now distil several key insights from the discussion on similar governance measures that are relevant for NeurIPS' broader impact requirement:
	
	\begin{itemize}
		\item Even for well-defined impact scopes (e.g. impacts on human subjects) there can be high variance in how impact assessments are evaluated \citep{dziak_variations_2005,larson_survey_2004,shah_how_2004,mcwilliams_problematic_2003,goldman_inconsistency_1982}. Review quality, consistency and process transparency are key desiderata \citep{whitney_principal_2008}, as is expertise by those who perform the evaluation \citep{bozeman_broad_2009,holbrook_peer_2011,bozeman_socio-economic_2017}. 
		\item If impact statements are evaluated by peers but community norms and best practice have not yet fully developed, disagreements may undermine efforts of ethical oversight \citep{kenneally_cyber-security_2014}.
		\item Researchers can feel overwhelmed by the task of having to consider broader impacts, or fail to consider them, if they do not receive appropriate training and guidance \citep{owen_responsible_2010}. 
		\item Assessment of broader impacts can benefit from the inclusion of specialists from other disciplines \citep{owen_responsible_2010}.
	\end{itemize}
	
	
	\section*{Effects of broader impact statement requirements}
	
	\subsection*{Potential positive effects}
	
	Introducing a broader impact requirement might have various potential positive effects. Building on and extending Hecht et al.'s assessment \citep{hecht_its_2018}, we identify four areas that could benefit in particular: \textbf{anticipation}, \textbf{action}, \textbf{reflection and awareness}, and \textbf{coordination}. These four categories closely resemble the EPSRC AREA framework on responsible research and innovation, which holds that science and innovation can better benefit society by following the steps: Anticipate, Reflect, Engage, and Act \citep{epsrc_anticipate_2020,owen_responsible_2012}. If the broader impact requirement proves to be successful, it could be seen as a significant contribution to the development of a responsible research practice. 
	
	We briefly outline the benefits within each category. 
	
	\textbf{Anticipation}. A thorough understanding of the societal effects of AI technologies should undergird all actions intended to improve its impacts. 
	AI researchers are well positioned to improve this understanding due to their technical expertise and in-depth knowledge of both their own research and the field more broadly. In particular, they are in an advantageous position to 
	
	\begin{enumerate}[(i)]
		\item anticipate potential applications of their research, and
		\item identify technical limitations and technical risks associated with such applications. 
	\end{enumerate}
	
	As such, impact statements can point towards opportunities, limitations, and risks that otherwise might not have gained attention \citep{stilgoe_developing_2013}. 
	
	Researchers writing impact statements may also lead to certain impacts being anticipated sooner. This can alleviate the Pacing Problem: the significant time lag between possibly disruptive innovation and adequate regulatory responses to many emerging technologies \citep{marchant_growing_2011,owen_responsible_2010}. The broader impact statements can help anticipate potential challenges at the research and development stage, allowing more time for identifying, evaluating, and ultimately acting to improve anticipated impacts. 
	
	\textbf{Action}. The anticipation of impacts helps stakeholders take appropriate action. For example, technical researchers can investigate mitigating solutions to potential negative consequences, or change the direction of their research so as to minimise negative impacts or to seize opportunities for positive impact \citep{hecht_its_2018}. Policymakers can enact policies to prevent rather than mitigate the impacts of technology after the fact. The public can pressure companies to release products that benefit society, and companies can better align their research with the public interest. 
	
	\textbf{Reflection and awareness}. By inviting researchers to reflect on the impacts of their own research, the requirement can contribute to raising awareness about issues associated with particular research or the field more generally, among both readers and authors. In time, this could produce a generation of researchers who have grappled with thinking about impacts and who chose to steer their research agendas towards opportunities to benefit society as a result. Reflection on impacts can furthermore help researchers to align their  research with the interests of other stakeholders, including the public, and in particular those most at risk of being adversely affected by downstream applications. 
	
	\textbf{Coordination}. The statements can be used to discover where there is consensus and disagreement on risks and opportunities within the AI community. Consensus on the existence of particularly risks can aid policy-making efforts. Disagreement can highlight areas where further dialogue is needed. In addition, the requirement may help build bridges with other communities in that it may help experts from other domains identify issues that need further investigation. There is also some evidence that requiring researchers to consider the broader impacts of their research can lead to an increase in outreach and cross-disciplinary collaboration \citep{owen_responsible_2010}.
	
	\subsection*{Potential negative effects}
	
	Drawing on the lessons from the previous case studies, and reflecting on the specifics of the NeurIPS requirement, we list the following potential negative outcomes.
	
	\textbf{Quality deficits}. Statements risk being uninformative, biased, misleading, or overly speculative. Quality deficits and ``lack of substance'' \citep[p.185]{bozeman_broad_2009} are issues that have also been flagged repeatedly in the context of NSF grant application impact statements and impact assessments more broadly \citep{frodeman_sciences_2007,gray_review_1999}. In the case of NeurIPS, several referees commented on social media about quality deficits of impact statements. 
	
	\textbf{Trivialisation of ethics and governance}:  Researchers might form the impression that it is possible to fully anticipate the ethical and societal consequences of one's research in such a statement, thereby trivialising the complexity of the task and the efforts needed \citep{holbrook_peer_2011,european_commission_assessing_2005,spaapen_introducing_2011}. Adequately anticipating impacts arguably requires expertise in the relevant ethical and social disciplines, good theory, and careful study of empirical evidence. Just as one might consider it inappropriate to ask ethicists or social scientists to include a statement on the existence of computationally feasible methods to implement concepts such as fairness, asking computer scientists to address the ethical and societal consequences of their work might not only lead to unsatisfactory results, but could contribute more widely to a trivialisation of ethics and governance itself. Additionally, there is a risk that shallow assessments become more prevalent than intellectually rigorous in-depth study. 
	
	\textbf{Negative attitudes}. Researchers might find writing the statements burdensome, confusing, punitive, or may perceive the activity to be of little value. In the context of NSF grant application impact statements, several scientists have expressed views along these lines \citep{tretkoff_nsfs_2007}. In a worse case scenario, researchers might develop negative attitudes towards responsible research and innovation more generally.
	
	\textbf{False sense of security}. Researchers (and readers) might feel a false sense of security if positive impacts are overstated, risks are underestimated, or if some risks are omitted altogether.
	
	\textbf{Unintended signalling}. The requirement or the statements themselves could signal messages that are unintended by the conference organisers, such as:
	\begin{enumerate}[(a)]
		\item \textit{The focus on individual researchers}: The requirement may signal that individual technical researchers are in the best position to reason about impact and make ethical judgements about their own work. In reality, the mitigation of harms requires a collective effort and input from many different stakeholders \citep{kenneally_cyber-security_2014}.
		\item \textit{Add-on}: The requirement could suggest to researchers that ethics is a tick-box exercise that is added to one's research as an afterthought. The UK Research and Innovation body removed their ``Pathways to Impact'' requirement for funding applications because they wanted impacts to be embedded in the project proposal rather than being a separate consideration \citep{uk_research_and_innovation_pathways_2020}.
	\end{enumerate}
	
	\textbf{Polarisation of the research community}. Polarisation of the research community on this topic may occur along political or institutional lines, or based on researcher's access to relevant resources. Similarly, scientific articles may become more politicised. 
	
	The above list illustrates how the introduction of broader impact requirements might miss its target (quality deficits) or lead to undesired effects on individual researchers (false sense of security, unintended signalling), the community (polarisation), or the field of AI ethics and governance as a whole (negative attitudes, trivialisation). Many of these also apply to discussions of \textit{positive impacts}, which are more routinely discussed in AI research papers. However, we believe the potential negative effects listed may become more acute when discussion of negative impacts are made compulsory, and when impact statements are treated as a separate part of the research.

	\section*{Causes and challenges}
	
	We now identify a number of causes for these potential negative effects, which in turn will allow us to develop a list of tentative suggestions for conference organisers and the community at large.

	\textbf{The complexity and difficulty of the task}. Foreseeing the downstream uses of AI research, especially foundational research, is notoriously difficult. This is particularly relevant for general purpose technologies like machine learning (ML), in which one technique may be applicable to a broad range of potential applications. Anticipating the societal impacts of those applications is even more challenging, even for social scientists and ethicists
	who have specialised in the task \citep{holbrook_peer_2011,european_commission_assessing_2005,spaapen_introducing_2011}. In contrast to social scientists, however, computer scientists typically have not received the relevant training that would allow them to address the complexity and nuances behind impact anticipation, nor can they be expected to be as familiar with the relevant literature.     
	
	\textbf{Lack of best practice and guidance}. There is currently no established best practice for writing impact statements. The experiment by Owen and Goldberg discussed above demonstrates the importance of guidance for researchers \citep{owen_responsible_2010}. Numerous studies that show high variance in IRB approval, as well as the mentioned ACM SIGCOMM 2015 incident, illustrate the importance of having clear guidelines and benchmarks that (a) help researchers to write adequate statements, and (b) allow for adequate and consistent assessments \citep{dziak_variations_2005,larson_survey_2004,shah_how_2004,mcwilliams_problematic_2003}. 
	
	\textbf{Lack of explanation of purpose.} Currently, little explanation has been provided regarding the \textit{precise} purpose and motivation of these statements. For example, whether the aim is to raise awareness, to anticipate impacts, to identify potentially unethical research, to encourage reflection or discussion, to motivate researchers to take responsibility, or any combination of the above. Such lack of clarity may dampen motivation, and lead to confusion. Clarity of purpose is also required to understand whether the initiative is meeting its aims.
	
	
	\textbf{Lack of procedural transparency and justice}. Transparency of procedures regarding the evaluation of impact statements, and how they relate to the ethical review process and acceptance decisions can help set expectations and create accountability for the process. Lack of transparency, on the other hand, as well as procedural injustice which causes decisions to be perceived as being influenced by bias, political opinions, or prejudice, may not only lead to inadequate evaluation, but also to growing resistance amongst researchers \citep{keith-spiegel_what_2006}.
	
	\textbf{High opportunity costs}. Writing a high quality impact statement takes time. Researchers routinely experience a substantial amount of pressure to deliver high-quality research articles under tight conference deadlines. If there is a trade-off between working on a research article and the broader impact statement, it is likely that researchers will opt for the first, especially when the broader impact statement does not substantially affect conference admission.
	
	\textbf{Institutional and social pressure}. Researchers may experience pressure---psychological, social, or institutional---to primarily emphasise the benefits of their work rather than potential downsides. Indeed, only emphasising the potential upsides of one's research is already standard practice \citep{hecht_its_2018}. As a result, researchers may be incentivised to focus on minor risks or those that do not threaten their own or their organisation's interests. Researchers working in a commercial setting might be particularly affected, as their employers have incentives to downplay certain risks to save commercial interests or prevent legal liabilities. It has been noted that corporations have an incentive to reframe ethics discussions in such a way that they do not challenge how the firm operates in any significant way \citep{bietti_ethics_2020}. This is of increasing significance as the number of papers from corporate affiliations increases over time \citep{hagendorff_big_2020}. 
	
	\textbf{Cognitive and social biases}. Cognitive and social biases may affect authors and referees of broader impact statements. Phenomena such as motivated reasoning may lead to gaps or belittlement of risks \citep{stanovich_myside_2013}. Several other biases, such as the framing effect \citep{plous_psychology_1993}, ambiguity effect \citep{curley_psychological_1986}, or confirmation bias \citep{nickerson_confirmation_1998}, may similarly affect the quality and the comprehensiveness of impact statements.

	\section*{Tentative suggestions}
	
	For conferences that choose to implement a broader impact requirement, we propose a number of tentative suggestions to help address the above challenges. We group these under \textbf{transparency}, \textbf{guidance}, \textbf{incentives} and \textbf{deliberation}, as summarised in Figure \ref{table2}. Many of the listed suggestions are not limited to broader impact statements, but apply to the peer-review process more generally. We emphasise that continuing or instigating a broader impact requirement should be contingent on the benefits outweighing the risks and costs, as revealed through (ongoing) deliberation.

	\begin{table}[ht] 
		\centering
		\def\arraystretch{1.5}
		\begin{tabular}{|r|l|}
			\hline
			\textbf{Transparency} &
			\begin{tabular}[c]{@{}l@{}}Purpose, motivation and expectations \\ Procedural transparency\\ Accountability mechanisms\end{tabular} \\ \hline
			\textbf{Guidance}   & \begin{tabular}[c]{@{}l@{}}Guidance   for researchers\\ Guidance   for reviewers \\ Communication   channels\end{tabular} \\ \hline
			\textbf{Incentives} & \begin{tabular}[c]{@{}l@{}}Peer-review \\ Outside expert involvement \\ Encouragement to cite\\ Prizes\end{tabular}  \\ \hline
			\textbf{Deliberation} &
			\begin{tabular}[c]{@{}l@{}}Forums   for deliberation   \\ Data-driven   deliberation    \\ Minimisation of reputational and legal   costs\end{tabular} \\ \hline
		\end{tabular}
		\caption{Tentative suggestions for the implementation of broader impact requirements.}
		\label{table2}
	\end{table}
	
	\subsection*{Improving transparency and setting expectations}
	
	Improving transparency and setting expectations can help address a number of the potential negative effects and significantly improve deliberation on both the purpose and implementation of broader impact requirements. 
	
	\textbf{Purpose, motivation and expectations}. Conference organisers need to communicate the purpose and motivations for introducing the broader impact statement requirement, as well as set expectations about length, format, and scope to reduce ambiguity. This can motivate and help researchers to better meet expectations, which in turn could make impact statements more consistent and of higher quality. 
	
	\textbf{Procedural transparency}. Researchers need visibility of how their impact statement is assessed and how this interacts with the peer-review process. In the case of NeurIPS, there still exists ambiguity regarding the review and assessment of impact statements. Procedural transparency furthermore allows for improved deliberation on the process by the community. 
	
	\textbf{Accountability mechanisms}. High variance in review results can be a problem for impact assessments \citep{larson_survey_2004,shah_how_2004,mcwilliams_problematic_2003,goldman_inconsistency_1982}. Established benchmarks against which statements are assessed are indispensable to improve reliability of outcomes and to create accountability. 
	
	\subsection*{Providing guidance}
	
	\textbf{Guidance for researchers}. In order to systematically and effectively address the broader impacts of their work, researchers need access to relevant guidelines, tools and training. There already  exist a number of freely available impact assessment tools, but it should be made clear to researchers, which, if any, to use \citep{responsible_research_and_innovation_toolkit_nodate}. 
	In addition, authors may benefit from guidance specifically aimed at computer science researchers and the NeurIPS requirement \citep{ai_ethics_lab_homepage_2020,ashurst_guide_2020,hecht_suggestions_2020}. To make impact statements more comparable and stimulate deliberations over best practice, we recommend that conference organisers (or sub-communities) deliberate on and develop their own set of tools and guidelines, including example statements. The community should deliberate on how prescriptive the guidance (and requirement) should be. Guidance that is too prescriptive could result in boilerplate responses; insufficient guidance could lead to poorly written statements \citep{owen_responsible_2010}. Providing researchers with training sessions, workshops, or tutorials organised by e.g. NeurIPS or third parties (such as IEEE or PAI) might also prove beneficial. 
	
	\textbf{Guidance for reviewers}. If impact statements are subject to (peer) review, it is important that those assessing them are qualified to do so and are informed about benchmarks and relevant assessment criteria. This is necessary to ensure quality and reliability of assessments. 
	
	\textbf{Creating communication channels}. Enabling researchers to engage with experts in other relevant fields may improve the scope and quality of impact statements and provide researchers with more support \citep{owen_responsible_2010}. For example, funding schemes directed at improving communication between research communities have shown to successfully enhance cross-disciplinary collaboration  \citep{porter_research_2012}. Events, such as workshops at ML conferences, can facilitate discussion on the impacts of AI research and improve deliberation both within and across disciplines. ML conferences can invite social science or ethics speakers to their conferences to improve awareness and dialogue. Finally, researchers might benefit from having access to a pool of AI ethics experts who have agreed to be available for feedback and discussion.
	
	\subsection*{Setting incentives}
	
	\textbf{Peer-review}. Integrating impact statements into the peer-review process is one way to ensure researchers take the requirement seriously. We encourage this step, though the challenges highlighted above should be addressed. There are significant resource considerations to address---with over 11,000 submissions to NeurIPS 2020, the peer-review process is already under strain. Implementation into the peer-review process could take different forms, ranging from referee comments, or unsatisfactory statements being grounds for rejection.
	
	\textbf{Outside expert involvement in the peer-review process}. If statements are subject to peer-review, the involvement of experts from relevant fields such as AI ethics and governance can allow for a more comprehensive assessment of impacts and deliberation on best practice. In the case of NeurIPS 2020, this was the case for papers that had been flagged for ethical concerns (and that had strong technical reviews). The programme chairs praised the high quality of outside expert assessments \citep{lin_what_2020}.
	
	\textbf{Encourage authors to cite relevant work and impact statements}. Researchers face strong incentives to ensure their work is widely cited. Tapping into this incentive, authors should be encouraged to cite relevant, high-quality impact statements as well as papers whose core research is positively influenced by societal considerations. 
	
	\textbf{Prizes}. Rewarding researchers for conscientious and well-written impact statements can encourage them to engage. Rewards could take different forms, including awards akin to best-paper awards. 
	
	\subsection*{Public and community deliberation}
	
	As we learn from experience of the initiative, conference organisers and the community need to deliberate publicly about its value, shortcomings, and future. 
	
	\textbf{Establish forums for deliberation}. Conference organisers should collect feedback from conference participants, publish their evaluation of the requirement, and reason publicly about the impact statement's value, shortcomings, and future. 
	
	\textbf{Provide relevant data}. Conference organisers or independent researchers should collect and publish data pertaining to the success and shortcomings of the requirement. Analysis could focus on (i) indicators of the quality of statements, (ii) whether institutional incentives appear to influence the considerations included, and (iii) the community's attitude towards the requirement. 
	
	\textbf{Investigate ways to minimise reputational or legal costs}. Researchers and their employers may fear legal or political backlash for highlighting certain risks, causing pressure to de-emphasise these risks. This needs to be addressed by the community and discussed in more detail with the stakeholders involved.

	We acknowledge that these suggestions will require additional resources from the ML community. However, some suggestions may reduce costs for individual researchers (such as improved guidance) and for many suggestions, the costs will reduce over time. Moreover, steps taken to encourage deliberation will not only benefit this initiative, but community-led governance initiatives more broadly. While many would argue that the responsibility to reduce harms justifies these costs, the community will need to continue to investigate the benefits and risks, and compare this initiative to other governance options.
	
	\section*{Conclusion}
	NeurIPS' novel governance initiative provides an important opportunity for the AI research community to reflect on its role and responsibilities in addressing societal impact. Our analysis shows that there are several challenges involved with the introduction of such broader impact requirements.  We do not believe, however, that these challenges are insurmountable. By offering a list of tentative suggestions we hope to have provided a starting point for further deliberation on these and similar governance initiatives.

	\section*{Acknowledgements}
	We thank Josh Tenenbaum, Yarin Gal, Toby Shevlane and colleagues at the Centre for the Governance of AI for helpful feedback and comments. 
	
	\section*{Competing interests}
	The authors declare no competing interests.


\end{document}